\documentclass[twocolumn,nofootinbib,aps,showpacs,floats,letterpaper,floatfix,groupedaddress]{revtex4}

\usepackage{graphicx,epsf, epsfig, amssymb}% Include figure files
\usepackage{bm}
\usepackage{longtable}

\def\be{\begin{equation}}
\def\ee{\end{equation}}
\def\beq{\begin{eqnarray}}
\def\eeq{\end{eqnarray}}

\newcommand{\secintro}{I}
\newcommand{\secsetup}{II}
\newcommand{\secresults}{III}
\newcommand{\secconclusion}{IV}

\begin{document}

\title{The high-energy collision of two black holes}

\author{Ulrich Sperhake$^{1}
%\footnote{Electronic address: ulrich.sperhake@uni-jena.de}
$, Vitor Cardoso$^{2,3}%\footnote{Electronic address: vcardoso@fisica.ist.utl.pt}
$, Frans Pretorius$^{4}%\footnote{Electronic address: fpretori@princeton.edu}
$, Emanuele Berti$^{5}%\footnote{Electronic address: berti@wugrav.wustl.edu}
$, Jos\'e A. Gonz{\'a}lez$^{6}%\footnote{Electronic address: jose.gonzalez@ifm.umich.mx}
$}
\affiliation{${^1}$ Theoretisch Physikalisches Institut, Friedrich Schiller Universit\"at, 07743 Jena, Germany}
\affiliation{${^2}$ %Centro Multidisciplinar de Astrof\'{\i}sica -
CENTRA, Departamento de F\'{\i}sica, Instituto
Superior T\'ecnico, Av. Rovisco Pais 1, 1049-001 Lisboa, Portugal}
\affiliation{${^3}$ Department of Physics and Astronomy, The University of
Mississippi, University, MS 38677-1848, USA}
\affiliation{${^4}$ Department of Physics, Princeton University, Princeton, NJ 08544, USA}
\affiliation{${^5}$ Jet Propulsion Laboratory, California Institute of Technology, Pasadena, CA 91109, USA}
\affiliation{${^6}$ Instituto de F\a'{\i}sica y Matem\a'aticas,
Universidad Michoacana de San Nicol\a'as de Hidalgo, Edificio C-3,
Ciudad Universitaria. C. P. 58040 Morelia, Michoac\a'an, M\a'exico}

%\date{\today}

\begin{abstract}

  We study the head-on collision of two highly boosted equal mass, nonrotating
  black holes. We determine the waveforms, radiated energies, and mode
  excitation in the center of mass frame for a variety of boosts.
  For the first time we are able to compare
  analytic calculations, black hole perturbation theory, and strong field,
  nonlinear numerical calculations for this problem. Extrapolation of our
  results, which include velocities of up to $0.94c$, indicate that in the
  ultra-relativistic regime about $14\pm 3 \%$ of the energy is converted into
  gravitational waves. This gives rise to a luminosity of order
  $10^{-2}c^5/G$, the largest known so far in a black hole merger.

\end{abstract}


%04.25.D-    Numerical relativity
%04.25.dc    Numerical studies of critical behavior, singularities, and cosmic censorship
%04.25.dg    Numerical studies of black holes and black-hole binaries
%04.50.-h    Higher-dimensional gravity and other theories of gravity
%04.50.Cd    Kaluza–Klein theories
%04.50.Gh    Higher-dimensional black holes, black strings, and related objects
%04.60.Cf    Gravitational aspects of string theory
%04.70.-s    Physics of black holes
%04.70.Bw    Classical black holes
%04.70.Dy    Quantum aspects of black holes, evaporation, thermodynamics
%04.80.-y    Experimental studies of gravity
%04.80.Cc    Experimental tests of gravitational theories
%11.25.Mj Compactification and four-dimensional models
%11.10.Kk    Field theories in dimensions other than four

\maketitle

%%%%%%%%%%%%%%%%%%%%%%%%%%%%%%%%%%%%%%%%%%%%%%%%%%%%%%%%%%%%%%%%%%%%%%%%%%%%%%
%\section{\label{sec:intro} Introduction}
\noindent{\bf{\em \secintro. Introduction.}}
%%%%%%%%%%%%%%%%%%%%%%%%%%%%%%%%%%%%%%%%%%%%%%%%%%%%%%%%%%%%%%%%%%%%%%%%%%%%%%
An important and long-standing problem in general relativity concerns the
ultra-relativistic scattering of black holes (BHs). This is one of the most
violent events one can conceive of in the theory.
%, where the nonlinear effects of strong field gravity are expected
% to play a dominant role.
The lack of
solutions has spurred much speculation about what may happen in this regime.
For example, these events are a natural testing ground for the cosmic
censorship conjecture: is there a class of initial conditions where they
generically lead to the formation of a naked singularity, or do event horizons
always form to cloth singular behavior in the geometry?

Related questions concern the ultra-relativistic scattering of particles.  If
the center of mass (CM) energy is beyond the Planck scale, gravity is expected
to dominate the interaction. Furthermore, since the kinetic energy dominates
over the rest mass energy, the gravitational interaction should be rather
insensitive to the structure of the particles, implying that the
trans-Planckian scattering of point particles should be well described by
BH scattering \cite{banks_fischler}.
This is of particular relevance for recent proposals to solve the hierarchy
problem by adding ``large'' extra dimensions \cite{Arkani-Hamed:1998rs}, or an
extra dimension with a {\em warp} factor \cite{Randall:1999ee}, thus producing
an effective electroweak Planck scale.  This offers the exciting possibility
that BHs could be produced in particle colliders and ultra high-energy
cosmic ray interactions with the atmosphere \cite{banks_fischler,bhprod}. A
naive estimate of the cross section for $M_{\rm Pl} \sim 1\,{\rm TeV}$
predicts that super-TeV particle colliders will produce BHs at a rate of a few
per second, making the Large Hadron Collider (LHC) at CERN a potential black
hole factory.  An important element to search for BH production
signatures is to understand the BH scattering process, and in
particular the energy lost to gravitational radiation. Given that the beam
commissioning to 7 TeV is scheduled for late 2008, this is a timely research
topic.  Further interesting applications of high-speed BH collisions
to high-energy physics have recently been suggested by the AdS/CFT
correspondence conjecture~\cite{adscft}.  Particularly intriguing is the
possibility of using this duality to understand properties of the quark-gluon
plasma formed in gold ion collisions at Brookhaven's Relativistic Heavy Ion
Collider (RHIC) through a study of ultra-relativistic BH collisions in
AdS~\cite{Nastase:2005rp}.

Early attempts to understand the ultra-relativistic BH scattering
problem were based on work by Penrose \cite{penrosetalk} in the 1970s.  He
modeled the spacetime metric as the union of two Aichelburg-Sexl waves
\cite{Aichelburg:1970dh}, describing the collision of two infinitely boosted
Schwarzschild BHs,
and found a closed trapped surface at the moment of collision, giving an
upper limit of roughly $29\%$ of the initial energy of the spacetime radiated
in gravitational waves.  Beyond the collision event the solution is
unknown. Given the extreme conditions of high-speed scattering it is unlikely
that analytic solutions describing the full dynamics of the spacetime will be
found, and therefore numerical methods must be employed. Only recently have
long-term stable numerical evolutions of black-hole binaries been achieved
\cite{Pretorius:2005gq}.  The flurry of subsequent activity exploring the
merger process has so far exclusively focused on rest-mass dominated scenarios
(see~\cite{Pretorius:2007nq} for a review).

In this {\em Letter} we report the first numerical solutions describing the
collision of two equal mass BHs in the regime where the initial energy of the
system is dominated by the kinetic energy of the BHs.  In Sec.~\secsetup\ we
describe the problem setup, including the numerical code and initial
conditions. We also review some existing analytical approximations to aspects
of the problem, which will be important both to interpret the numerical
results and to give some confidence in extrapolations of the results to
infinite boost.  In Sec.~\secresults\ we present the primary results, focusing
on the gravitational waves emitted during the collision.  Concluding remarks
are given in Sec.~\secconclusion.  Unless stated otherwise, we use geometrical
units $G=c=1$.

%%%%%%%%%%%%%%%%%%%%%%%%%%%%%%%%%%%%%%%%%%%%%%%%%%%%%%%%%%%%%%%%%%%%%%%%%%%%%%
%\subsection{\label{sec:analytic} Comparison with other methods}
\noindent{\bf{\em \secsetup. Numerical Setup and Analysis Tools.}}
%%%%%%%%%%%%%%%%%%%%%%%%%%%%%%%%%%%%%%%%%%%%%%%%%%%%%%%%%%%%%%%%%%%%%%%%%%%%%%
%%%%%%%%%%%%%%%%%%%%%%%%%%%%%%%%%%%%%%%%%%%%%%%%%%%%%%%%%%%%%%%%%%%%%%%%%%%%%%
%\section{\label{sec:setup} Numerical setup and procedure}
%%%%%%%%%%%%%%%%%%%%%%%%%%%%%%%%%%%%%%%%%%%%%%%%%%%%%%%%%%%%%%%%%%%%%%%%%%%%%%
The numerical simulations presented here have been performed with the {\sc
  Lean} code, described in detail in~\cite{Sperhake:2006cy}, where head-on
collisions of different classes of initial data were compared. Here we
exclusively study evolutions of puncture initial data \cite{Brandt:1997tf}
describing two equal mass, non-spinning, boosted BHs colliding with zero
impact parameter in the CM frame.  The initial coordinate separation between
the punctures is set to $r_0$, and the boosts are prescribed in the form of
non-vanishing Bowen-York \cite{Bowen:1980yu} parameters $\pm P$ for the
initial linear momentum of either BH. The Hamiltonian constraint is solved
using Ansorg's spectral solver {\sc TwoPunctures} \cite{Ansorg:2004ds}.
The irreducible masses $M_{\rm irr1,2}$ of the BHs are estimated from their
apparent horizon areas, calculated using Thornburg's apparent horizon finder
{\sc AHFinderDirect} \cite{Thornburg:1995cp}. This enables us to calculate the
BH masses $M_{1,2}$ from Christodoulou's
\cite{Christodoulou1970}
relation $M_{1,2}^2 = M_{{\rm irr1,2}}^2+ P^2$,
  from which we define the Lorentz boost parameter $\gamma \equiv
  M_{1,2}/M_{\rm irr1,2}$ (cf.~\cite{Cook1989}).
From a numerical point of view, simulations with large values of $\gamma$ are
challenging, partly because the Lorentz contraction decreases the
smallest length scale that needs to be resolved.
Thus mesh-refinement is essential, and here it is
provided via the {\sc Carpet} package \cite{Carpetweb}.

We use the Newman-Penrose scalar $\Psi_4$ to measure gravitational radiation.
At an extraction radius $r$ from the center of the collision we decompose
$\Psi_4$ into multipole modes $\psi_{lm}$ of the spherical harmonics of
spin-weight $-2$, ${_{-2}}Y_{lm}$, according to
$\Psi_4(t,r,\theta,\phi)=\sum_{l=2}^\infty \sum_{m=-l}^l
\,{_{-2}}Y_{lm}(\theta\,,\phi)\, \psi_{lm}(t,r)$.
Due to the symmetries of this problem, the only non-vanishing multipoles all
have even $l$, $m=0$, and are purely real, corresponding to a single
polarization state $h_+$.  The energy spectrum and luminosity of the radiation
are given by
\beq \frac{dE}{d\omega}&=&\sum_l
\frac{1}{16\pi^2}\frac{|\hat{\psi}_{l0}(\omega)|^2}{\omega^2}
\equiv \sum_l \frac{dE_l}{d\omega}\,,\label{spectrum} \\
\frac{dE}{dt}&=&\sum_l
\lim_{r\rightarrow \infty}\frac{r^2}{16\pi} \left|
\int_{-\infty}^{t} \psi_{l0}(\tilde{t}) d\tilde{t} \right|^2
\equiv \sum_l \frac{dE_l}{dt}\,,
\label{luminosity}\eeq
respectively, where a hat denotes Fourier transform.

Our results are affected by three main sources of uncertainties: finite
extraction radius, discretization and spurious initial radiation. We reduce
the error arising from finite extraction radius by measuring the waveform
components at several radii, and fitting them to an expression of the form
$\psi_{lm}(r,t)=\psi_{lm}^{(0)}(t)+\psi_{lm}^{(1)}(t)/r$.
The waveform ``at infinity'' $\psi_{lm}^{(0)}(t)$ is the quantity reported
throughout this work and used to calculate related quantities, such as the
radiated energy. The uncertainty in this extrapolated value is estimated by
performing a second fit including also a quadratic term $\psi_{lm}^{(2)}/r^2$,
and taking the difference between the first- and second-order fits. The
resulting uncertainty in the radiated energy is typically $\sim 3-5~\%$.

To estimate discretization errors we evolved the most challenging simulation
with $\gamma\approx 3$ with resolutions $h=M/174$, $M/209$ and $M/244$, where
$M=M_1+M_2$.
We observe convergence slightly below second order in the total radiated
energy, and use a conservative estimate of $10\%$ for the resulting error near
$\gamma\approx 3$, which drops to a few percent in the non-boosted case
(cf.~\cite{Sperhake:2006cy}).

Finally, the conformally flat puncture initial data is known to contain
spurious gravitational radiation,
% \cite{Cook2000},
which increases strongly with boost $\gamma$ (from a few times $10^{-5}$ for
BHs at rest, to about $8\%$ of the total ADM mass of the system for
$\gamma\approx 3$). In order to extract physically meaningful information, one
has to separate the spurious radiation from the radiation generated by the
collision itself. This is done by ``waiting'' for the spurious radiation to
pass the last extraction radius, and then discarding the earlier part of the
wave signal.
%\cite{Sperhake:2006cy}.
For large boosts, the amount of time between the trailing edge of the spurious
radiation and the leading edge of the waves emitted during the collision is
roughly $r_0/(4\gamma^2)$. Thus, the initial separation required to cleanly
extract the emitted signal increases rapidly with $\gamma$. Because large
separations require larger computational domains and longer run-times, the
spurious radiation effectively limits our ability to study very large
$\gamma$. With current resources, we were able to use initial separations of
up to $66M$ for $\gamma>2$, leading to an uncertainty
in the total radiated energy which  grows rapidly with boost,
reaching a value of $5\%$ for $\gamma \approx 3$.
By combining all errors, we estimate the total uncertainty in
the radiated energy to be about $15\%$ for $\gamma\approx 3$, about $10\%$
near $\gamma=2$ and a few percent for simulations with small velocities.
% We
%generally find the uncertainties to cause an underestimation of the radiated
%energy, and thus believe the correct values to be closer to the upper ranges of
%our error bars.

High-energy collisions are uncharted territory for numerical relativity. It is
helpful, therefore, to have alternative methods for guidance and consistency
checks. Besides Penrose's bound, we will make extensive use of
extrapolations of Smarr's ``zero-frequency limit'' (ZFL) \cite{Smarr:1977fy}
and of point particle (PP) calculations \cite{Davis:1971gg},
where one considers a small object of mass $m$ colliding with a massive BH of
mass $M_{\rm BH}$ to linear order in $m/M_{\rm BH}$.

%%%%%%%%%%%%%%%%%%%%%%%%%%%%%%%%%%%%%%%%%%%%%%%%%%%%%%%%%%%%%%%%%%%%%%%%%%%%%%
%\subsection{\label{sec:curvature} Horizon deformation and properties}
\noindent{\bf{\em \secresults. Results.}}
%%%%%%%%%%%%%%%%%%%%%%%%%%%%%%%%%%%%%%%%%%%%%%%%%%%%%%%%%%%%%%%%%%%%%%%%%%%%%%
We ran a series of simulations from $\gamma=1$ to $\gamma\approx 3$, with
initial separations as discussed in the previous section. In all cases the
collision results in a single BH plus gravitational radiation, i.e. there is
no sign of any violation of cosmic censorship.  The final BH is born highly
distorted. We measure the distortion by taking the ratio ${\cal C}$ of the
proper equatorial to polar circumferences of the common apparent horizon
(CAH). For the range of boosts studied here, the peak value is well fitted by
the relation
${\cal C}^{\rm peak} \sim 1.5-0.5/\gamma$. Thus in the large-$\gamma$ limit
${\cal C}^{\rm peak}\sim 1.5$, in agreement with Penrose's result of ${\cal
  C}=\pi/2$ for a CAH consisting of two flat disks. After birth, the BH
settles down to a Schwarzschild solution, and the gravitational radiation can
be described as a superposition of quasinormal modes (QNMs) of the resulting
BH.

%%%%%%%%%%%%%%%%%%%%%%%%%%%%%%%%%%%%%%%%%%%%%%%%%%%%%%%%%%%%%%%%%%%%%%%%%%%%%%
%\subsection{\label{sec:waves} Waveforms}
%%%%%%%%%%%%%%%%%%%%%%%%%%%%%%%%%%%%%%%%%%%%%%%%%%%%%%%%%%%%%%%%%%%%%%%%%%%%%%
%
\begin{figure}[t]
\begin{center}
\begin{tabular}{c}
\epsfig{file=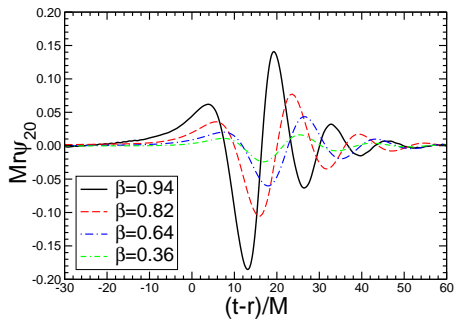,width=4.5cm,angle=-90} %&
\end{tabular}
\caption{Dominant multipolar component $\psi_{20}(t-r)$ for
different values of $\beta$, as indicated in the inset.
\label{fig:wavespsih2}}
\end{center}
\end{figure}
In Fig.~\ref{fig:wavespsih2} we show the dominant component $\psi_{20}$ of the
waveform from collisions with $\gamma=1.07,~1.3,~1.7,~3.0$ (corresponding to
$\beta=v/c\simeq 0.36,~0.64,~0.82,~0.94$, respectively). The origin of the
$(t-r)$ axis roughly corresponds to the instant of formation of a CAH. One can
identify three main parts in the waveforms: a precursor, a main burst at the
onset of the CAH formation and the final ringdown tail.  These seem to be
universal properties of collisions involving BHs and were observed in the past
in different settings \cite{Davis:1971gg,Anninos:1993zj}. The start of
ringdown, roughly associated with the absolute maxima $| \psi_{20}^{\rm
  peak}|$ in $|\psi_{20}|$, occurs $\sim 15M$ after the CAH formation,
independently of $\gamma$. Except for a small neighborhood around $\gamma \sim
1$, the maximal wave amplitude $|\psi_{20}^{\rm peak}|$ increases
monotonically with the boost factor.  The small dip in the wave amplitude for
small, but non-zero velocities has been seen before both in numerical
simulations and analytic predictions \cite{Baker:1996bt}. For moderate boosts,
we observe the absolute maxima in $\psi_{20}$ to be well approximated by
$|Mr\psi_{20}^{\rm peak}| \approx 0.26 + 0.48 \gamma^{-2}
      \left[1/4 +\log (1/2\gamma) \right]$
[cf.~Eq.~(\ref{etotic}) below].
The peak amplitude in the waveform $h_{20}$ is roughly $h_{20}^{\rm peak} \sim
\psi_{20}^{\rm peak}/\omega_{\rm QNM}^2$, where $\omega_{\rm QNM}$ is the
lowest ringdown frequency for the mode \cite{Berti:2005ys}.

\begin{figure}[t]
%\begin{center}
\begin{tabular}{c}
\epsfig{file=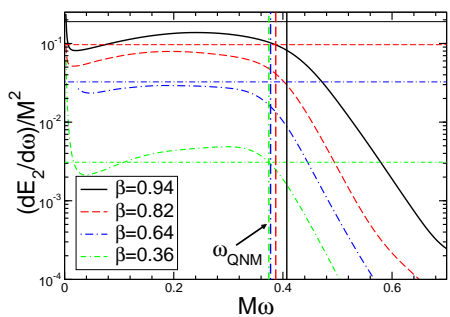,width=4.5cm,angle=-90} %&
\end{tabular}
\caption{Energy spectrum for $l=2$ and different values of $\beta$. Horizontal
  lines are the corresponding ZFL-PP predictions, vertical lines are the QNM
  frequencies of the final BH.
  \label{fig:spectra}}
%\end{center}
\end{figure}
Figure \ref{fig:spectra} shows the energy spectrum (\ref{spectrum}) for
collisions with different CM energy. For large CM energies, the spectrum is
nearly flat up to some cutoff frequency. A flat spectrum is predicted by the
ZFL and PP approaches, as indicated by the dotted lines in the figure.  The
cutoff frequency is well approximated by the least-damped QNM of the final
hole, marked by a vertical line. The spectrum increases at small frequencies
because of initial data contamination and finite-distance effects.

Our numerical results indicate that the peak luminosity (\ref{luminosity}) is
attained approximately $10M$ after the CAH formation. The peak luminosity is
about $5\times 10^{-3}$ for $\beta=0.9$, and may be as large as $10^{-2}$ as
$\gamma \to \infty$. Restoring units, we get $10^{-2}c^5/G\sim
3.6\times10^{57}{\rm erg}\, {\rm s}^{-1}$, the largest luminosity from a BH
merger known to date. This is two orders of magnitude larger than for the
infall from rest of two equal mass BHs, and one order of magnitude larger than
for the inspiral of equal mass binaries. Nevertheless, it is still two orders
of magnitude below the universal limit suggested by Dyson, $dE/dt \lesssim 1$
\cite{dyson}.

\begin{figure}[tb]
\begin{center}
\epsfig{file=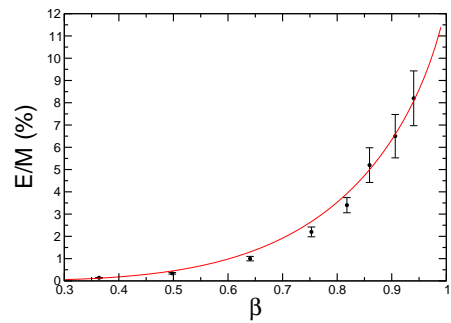,width=4.5cm,angle=-90} %&
%\begin{tabular}{c}
%\epsfig{file=PLOTS/energy.eps,width=6cm,height=4cm,angle=0}
%\end{tabular}
\caption{Total radiated energy (including error bars) as a function of
  $\beta$, and best fit using the ZFL prediction. \label{fig:envsfit}}
\end{center}
\end{figure}

The total energy $E$ radiated as a function of boost parameter is shown in
Fig.~\ref{fig:envsfit}. Error bars on the radiated energies are determined as
described in Sec.~\secsetup. We have verified that $E$ calculated from the
radiation (\ref{luminosity}) is consistent with alternative estimates obtained
by directly measuring the mass of the final hole from the CAH properties, and
by using the ringdown frequency to estimate the mass of the final hole
\cite{Berti:2005ys}.
The ZFL predicts the following functional form for the total radiated energy
as a function of CM~boost~$\gamma$:
\beq \frac{E}{M}=E_{\infty}\left(\frac{1+2\gamma^2}{2
\gamma^2}+\frac{(1-4\gamma^2)\log{(\gamma+\sqrt{\gamma^2-1})}}{2
\gamma^3\sqrt{\gamma^2-1}}\right)\,.
%\nonumber\\
%&\equiv&E_{\infty}\zeta
\label{etotic}\eeq
The quantity $E_{\infty}$ is some unknown cutoff parameter, which is also the
total fraction of energy radiated as $\gamma\rightarrow \infty$. By fitting
Eq.~(\ref{etotic}) to the numerical data we obtain
$E_{\infty}=0.14\pm0.03$. The ZFL is a perturbative calculation about
$\omega=0$, and its validity for our scattering problem is not
obvious. However, given the good agreement with our numerical results in the
kinetic-energy dominated regime $\gamma>2$, the extrapolation procedure should
provide a reasonably accurate estimate for $E_\infty$.

With regard to the multipolar contributions of the radiated energy,
we find that $E_4$ is at least one order of magnitude smaller than $E_2$ for
slow-motion collisions. This observation is consistent with the PP results for
an infall from rest \cite{Davis:1971gg}, which predict an exponential decrease
of $E_l$ with $l$.  For larger boosts the ZFL and PP approach predict a strong
increase in the relative contribution of higher multipoles, with $E_{l}\sim
M/l^2$ as $\gamma\rightarrow \infty$. Our numerical results are in reasonable
agreement with these calculations, as demonstrated in Table
\ref{tab:energyratio}.
\begin{table}[t]
  \centering \caption{\label{tab:energyratio} Relative multipolar contribution
    (in \%) and, in parentheses, the ZFL prediction.}
\begin{tabular}{ccccccc}
\hline
$\beta$ & 0.64& 0.75 & 0.82 &0.86 &0.90& 0.94    \\
\hline
$E_4/E_2$  &1.0(1.4)&2.4(3.4)&3.9(5.4)&5.0(7.3)&7.3(10)&11(14) \\
$10E_6/E_2$ &0.2(0.3)&1.1(1.5)&2.1(4.0)&4.2(7.5)&11(16)&33(30)\\
\hline \hline
\end{tabular}
\end{table}
The discrepancies still present are due to the relatively large uncertainties
in the energy carried by higher multipoles and to the breakdown of the ZFL
prediction for small boosts.

%%%%%%%%%%%%%%%%%%%%%%%%%%%%%%%%%%%%%%%%%%%%%%%%%%%%%%%%%%%%%%%%%%%%%%%%%%%%%%
%\section{\label{concl} Conclusions}
\noindent{\bf{\em \secconclusion. Conclusions.}}
%%%%%%%%%%%%%%%%%%%%%%%%%%%%%%%%%%%%%%%%%%%%%%%%%%%%%%%%%%%%%%%%%%%%%%%%%%%%%%
In 1971, Hawking \cite{hawking} placed an upper limit of $29\%$ on
the total energy radiated when two BHs, initially at rest,
coalesce. Numerical simulations of Einstein's equations
\cite{Anninos:1993zj} later showed that the true value is around
$0.1$\%---two orders of magnitude smaller than Hawking's bound.
Using a similar area theorem argument, Penrose \cite{penrosetalk}
derived an upper bound of $29\%$ for ultra-relativistic head-on
collisions (that the numerical values of the two bounds agree is
apparently just a coincidence). Here we have presented results
indicating that the answer in the high-energy limit is $0.14\pm
0.03$, slightly less than a factor of $2$ of Penrose's bound,
though quite close to the estimate of D'Eath and Payne computed
using perturbative techniques~\cite{D'Eath:1992qu}. Even though
our calculations are in 4D, a consequence of this to searches for
BH formation at the LHC is a warning that estimates of the
``missing energy'' based upon trapped surface calculations could
significantly overestimate this effect.

This long overdue study represents an important step towards a
full understanding of high-energy BH collisions. More accurate
evolutions using significantly larger boosts are mainly
inhibited by the junk radiation in the initial data.
More work is also
needed to study scattering with non-zero impact parameter, unequal
masses and non-zero spins. For applications to LHC and RHIC
physics, including the effects of extra dimensions, charge and AdS
asymptotics (for RHIC) will be necessary.

%%%%%%%%%%%%%%%%%%%%%%%%%%%%%%%%%%%%%%%%%%%%%%%%%%%%%%%%%%%%%%%%%%%%%%%%%%%%%%
%\section{Acknowledgements}
{\bf \em Acknowledgements.}
%%%%%%%%%%%%%%%%%%%%%%%%%%%%%%%%%%%%%%%%%%%%%%%%%%%%%%%%%%%%%%%%%%%%%%%%%%%%%%
This work was supported in part by DFG grant SFB/TR~7, by FCT - Portugal
through projects PTDC/FIS/64175/2006 and POCI/FP/81915/2007, and by the
Fulbright Foundation (V.C.). E.B. was supported by the NASA Postdoctoral Program
at JPL/Caltech, administered by Oak Ridge Associated Universities through a
contract with NASA. F.P. was supported by the Alfred
P. Sloan Foundation and NSF PHY-0745779.
Computations were performed at LRZ Munich, Milipeia at CFC in
Coimbra, and the Woodhen cluster at Princeton University.

\end{document}